\documentclass{webofc}
\usepackage[varg]{txfonts}   
\begin{document}
\title{Machine Learning as a Service for HEP}
%
%

\author{\firstname{Valentin} \lastname{Kuznetsov}\inst{1}\fnsep\thanks{\email{vkuznet@gmail.com}}}

\institute{Cornell University, Ithaca, NY, USA 14850}

\abstract{%
    Machine Learning (ML) will play significant role in success of the upcoming
    High-Luminosity LHC (HL-LHC) program at CERN. The unprecedented amount of
    data at the Exa-Byte scale to be collected by the CERN experiments in next decade
    will require a novel approaches to train and use ML models. In this
    paper we discuss Machine Learning as a Service (MLaaS) model which is
    capable to read HEP data in their native ROOT data format, rely on the
    World-Wide LHC Grid (WLCG) infrastructure for remote data access, and serve a
    pre-trained model via HTTP protocol. Such modular design opens up a
    possibility to train data at large scale by reading ROOT files from remote storages,
    avoiding data-transformation to flatten data formats currently used by ML
    frameworks, and easily access pre-trained ML models in existing infrastructure
    and applications.
}
\maketitle

\section{Introduction}

With the CERN LHC program underway, we started seeing an exponential acceleration of
data growths in High-Energy Physics (HEP) field.  By the end of Run II, the CERN
experiments were already operating in the petabyte (PB) regime, producing
$O$(100)PB of data each year. And, the new HL-LHC program will bring us to the
Exa-Byte scale. The usage of Machine Learning in the HEP is on the rise too.
It has been successfully used in online, offline reconstruction programs, and
there is huge interest to apply it for detector simulation, object
reconstruction, identification, MC generation, and beyond \cite{MLCWP}. But the
main obstacle of using ML frameworks and brining CS expertise in ML to HEP lies
in differences of data-formats used by ML practitioners and HEP users. In
particular, the former mostly rely on flat-format data representation, e.g.
CSV or NumPy data formats, while HEP data are stored in
tree based data-structures used by ROOT \cite{ROOT} data-format. As was pointed
out in HEP ML Community White Paper \cite{MLCWP}, the usage of ROOT data-format
outside of HEP practically does not exists. This fact creates an artificial gap
between ML and HEP communities.  The recent kaggle challenges, e.g. ATLAS for
identification of Higgs boson \cite{kaggleATLAS} and the cross-experiment tracking
ML challenge \cite{kaggletracking}, were specifically adopted (in terms of
input datasets) and presented to ML competitors in CSV data format. But, within
the HEP community these datasets are easily accessible, without any pre-processing
or transformation in the ROOT data-format. To close this gap, we present in this
paper a novel approach to use HEP ROOT data natively for training purposes,
reading ROOT files from remote storages via XrootD, and presenting pre-trained
models as a service accessible via HTTP protocol. Such Machine Learning as a
Service (MLaaS) modular design opens up a possibility to train ML models on PB
datasets remotely accessible from the Worlwide LHC Computing GRID (WLCG) sites
without requiring data transformation and data locality.

\section{Related work and solutions}
Machine Learning as a Service is a well known concept in industry, and major IT
companies offer these solutions to their customers. For example, Amazon ML,
Microsoft Azure ML Studio, Google Prediction API and ML engine, and IBM Watson
are good examples of MLaaS, see \cite{MLaaScomparison}.  Usually, MLaaS is used
an umbrella of various ML tasks such as data pre-processing, model training and
evaluation, and prediction results can be accessed by clients through REST
APIs. Even though they can provide very good results and interfaces, most of
the time these services are designed to cover standard use-cases.  For
instance, data are expected to be fed in flat based data formats. All data
preprocessing operations are performed automatically where a concrete service
identifies which fields are categorical and which are numerical, and it does not
ask a user to choose the methods of further data preprocessing.  The model
predictions are limited to well-established patterns, such as binary
classifications, multi-class classifications, and regressions.  Although quite
often MLaaS service providers offer pre-defined models that can be used to
cover standard use-cases, e.g. image classifications, etc. Obviously, all
them are designed to make a profit by charging customers on the amount of
predictions they want to make, or use tiered structure on the amount of calls
placed by clients. 

In HEP, usage of these services is quite limited though for several reasons.
Among them, the HEP ROOT data-format can't be used directly in any of these
services, and pre-processing operations may be more complex than offered by
service providers. For instance, the two HEP kaggle challenges
\cite{kaggleATLAS, kaggletracking} used custom HEP metrics for evaluation
procedure which is not available in out-of-the box industry solutions, and ML
workflow in both competitions is far from trivial, e.g the pre-processing step
required writing custom code to include event selection and perform other
steps. Therefore, after rounds of evaluations we found that provided solutions
most often are ineffective for HEP use-cases (cost and functionality-wise), even though
the CERN OpenLab initiative and others continue close cooperation with almost
all aforementioned service providers.

At the same time, various R\&D activities within HEP is underway.  For example:
hls4ml project \cite{hls4ml} targets ML inference in FPGAs, while SonicCMS
project \cite{SonicCMS} is designed as Services for Optimal Network Inference
on Coprocessors.  Both are designed for optimization of inference phase rather
than targeting the whole ML pipeline from reading data, to training and serving
predictions. At the moment we are unaware of any final product which can be
used as MLaaS in HEP. The novelty of the proposed solution is three fold.
First, we are proposing to use HEP ROOT files directly, either using them
locally or remotely, without requiring data transformation operations to flat
data format. Second, the training layer can use external 3rd party ML
frameworks, from well established ML, e.g. scikit-learn, libraries to
Deep-Learning (DL) frameworks such as TensorFlow, PyTorch and others. Third the
inference phase is provided via RESTful APIs of TensorFlow as a Service (TFaaS)
similar to industry solutions.  The latter does not require significant changes
of existing HEP infrastructures, frameworks and applications due to usage of
HTTP protocol between clients and TFaaS server(s).

\section{MLaaS architecture}\label{Architecture}
A typical ML workflow consists of several steps: acquire the data necessary
for training, use ML framework to train the model, and utilize the trained model for
predictions. This workflow can be further abstracted as data streaming,
data training, and inference phases. Each of these steps can be either
tightly integrated into application design or composed and used individually.
The choice is mostly driven by particular use cases. In HEP we can
define these layers as following, see Fig. \ref{fig:MLaaSArchitecture}:
\begin{figure}
\centering
\includegraphics[width=0.8\textwidth]{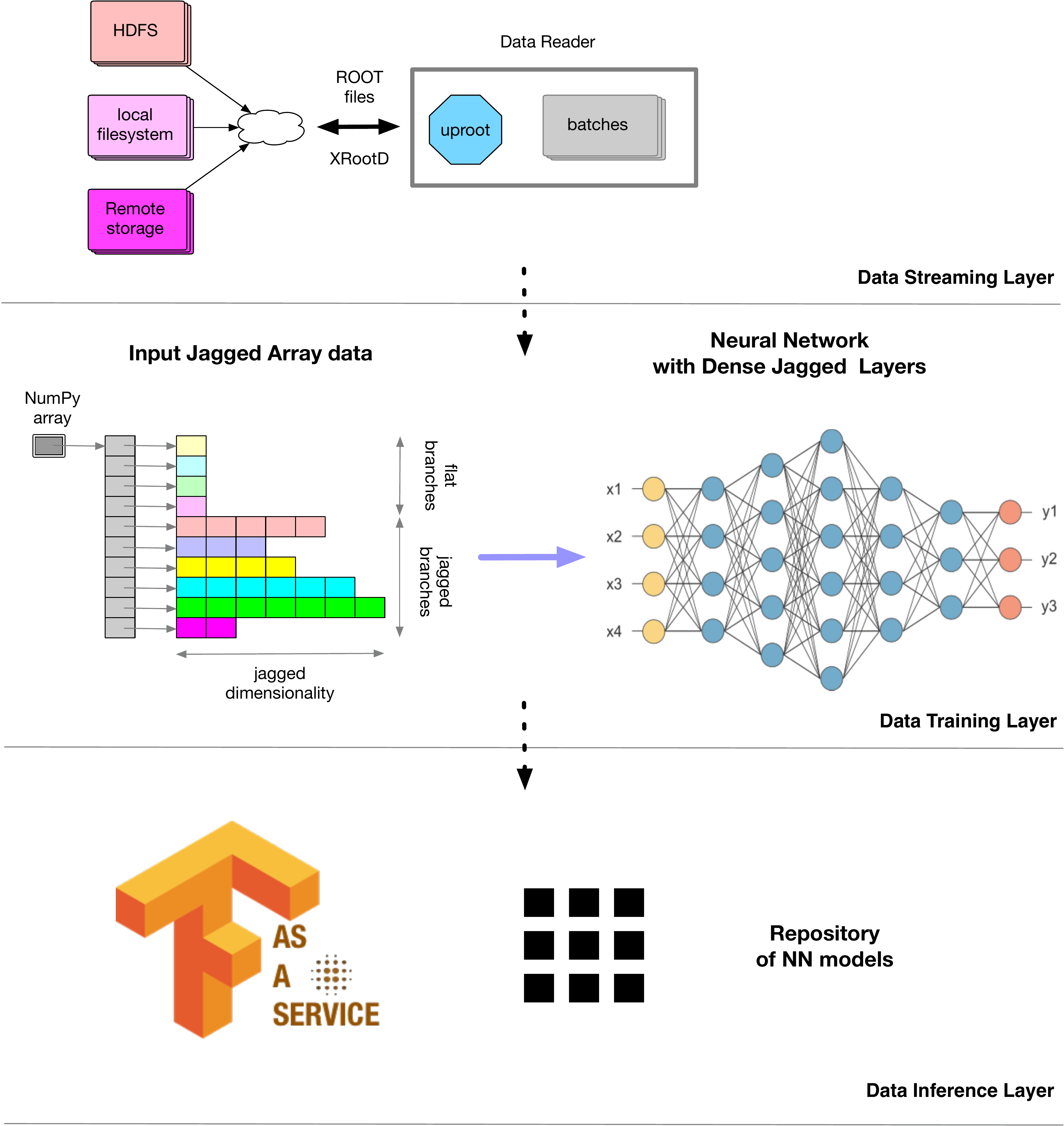}
    \caption{MLaaS architecture diagram representing three independent layers:
    data streaming layer to read local or remote ROOT files, a data training layer
    to feed Tree based HEP data into ML framework, and data inference layer
    via TensorFlow as a Service.}
\label{fig:MLaaSArchitecture}
\end{figure}
\begin{itemize}
\item {\bf Data Streaming Layer} is responsible for reading local and/or remote ROOT files,
    and streaming data batches upstream to the Data Training Layer. The implementation of
    this layer requires ROOT I/O layer with support of remote I/O file access;
\item {\bf Data Training Layer} represents a thin wrapper around standard ML libraries
    such as TensorFlow, PyTorch, and others. It reads data from the Data
    Streaming Layer in chunks, transforms them from ROOT TTree based
    representation to the format suitable for underlying ML framework and
    uses it for training purposes;
\item {\bf Data Inference Layer} refers to the inference part of pre-trained models and
    can be either tightly integrated within underlying HEP framework or represented
    as a Service (aaS).
\end{itemize}
Even though the implementation of these layers can differ from one experiment
to another (or other scientific domains/fields using ROOT files), it can be
easily generalized and be part of the foundation for generic MLaaS framework.
Further, we will discuss individual layers and outline particular sets of
problems which should be addressed in their implementation.

\subsection{Data Streaming Layer}\label{Streaming}
The data streaming layer represents a simple task of streaming data from local
or remote data storages. Originally reading ROOT files was mostly possible from
C++ frameworks, but recent development of ROOT I/O now allows to easily access
ROOT data from Python, and use XrootD protocol for remote file access. The main
development was done in uproot \cite{uproot} framework backed by the DIANA-HEP
initiative \cite{DIANAHEP}. The uproot library uses NumPy \cite{NumPy} calls
to rapidly cast data blocks in the ROOT file as NumPy arrays,
and provides integration with the XrootD protocol
\cite{XrootD}. Among the implemented features it allows a partial reading of
ROOT TBranches, non-flat TTrees, non TTrees histograms and more. It relies on
data caching and parallel processing to achieve high throughput. In our
benchmarks we were able to read HEP events at the level of $\sim
O(100)-O(1000)$kHz from local and from remote
storages\footnote{Speed varies based on many factors, including caching, type of storage and network bandwidth.}.

In our implementation of MLaaS, see Sect. \ref{Prototype}, this layer was
composed as a Data Generator which is capable of reading either local or remote
file(s) with a pre-defined size.  The batch data size can be easily fine tuned
based on the complexity of the event and available bandwidth.  The output of
the Data Generator was a NumPy array with flat and Jagged Array attributes, see
next Section for further discussion.

\subsection{Data Training Layer}\label{Training}
\begin{figure}
\centering
    \includegraphics[width=0.4\textwidth]{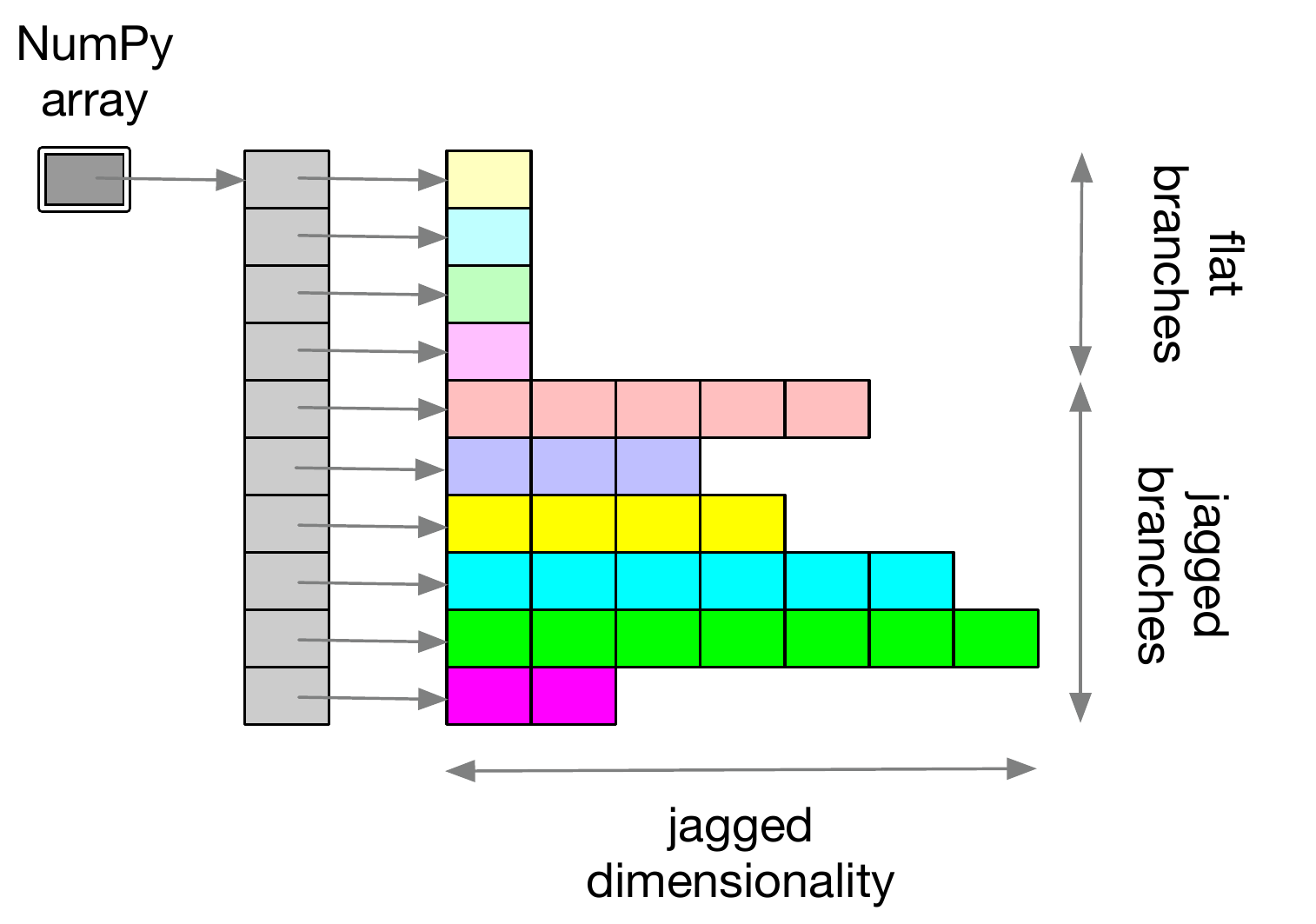}
    \caption{Jagged Array data representation. It consists of flat attributes followed by
    Jagged attributes whose dimensions vary event by event.}
\label{fig:JaggedArray}
\end{figure}

\begin{figure}
\centering
    \includegraphics[width=0.6\textwidth]{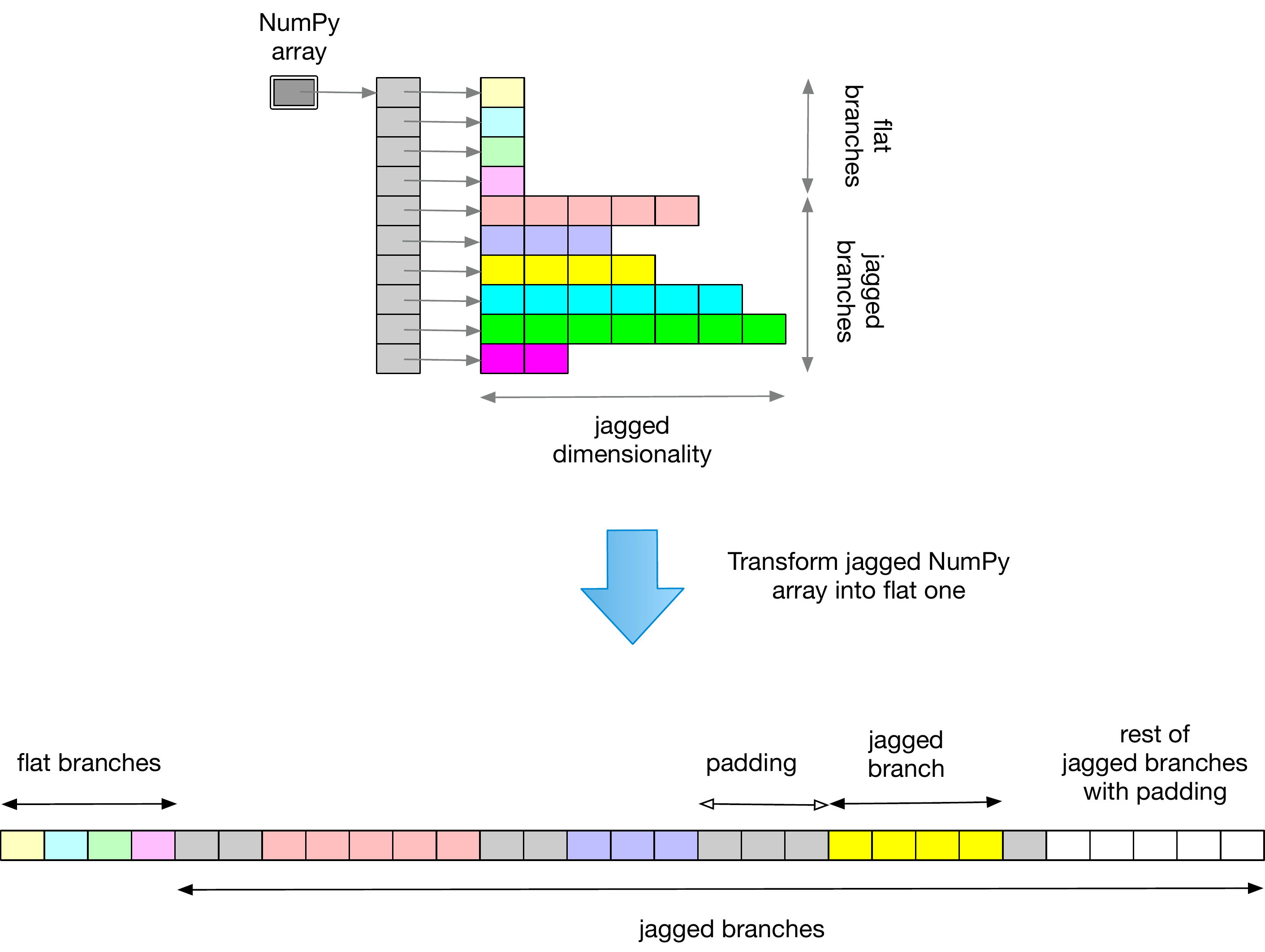}
    \caption{A vector representation of Jagged Array with padded values.}
\label{fig:JaggedArray2Vector}
\end{figure}

\begin{figure}
\centering
    \includegraphics[width=0.6\textwidth]{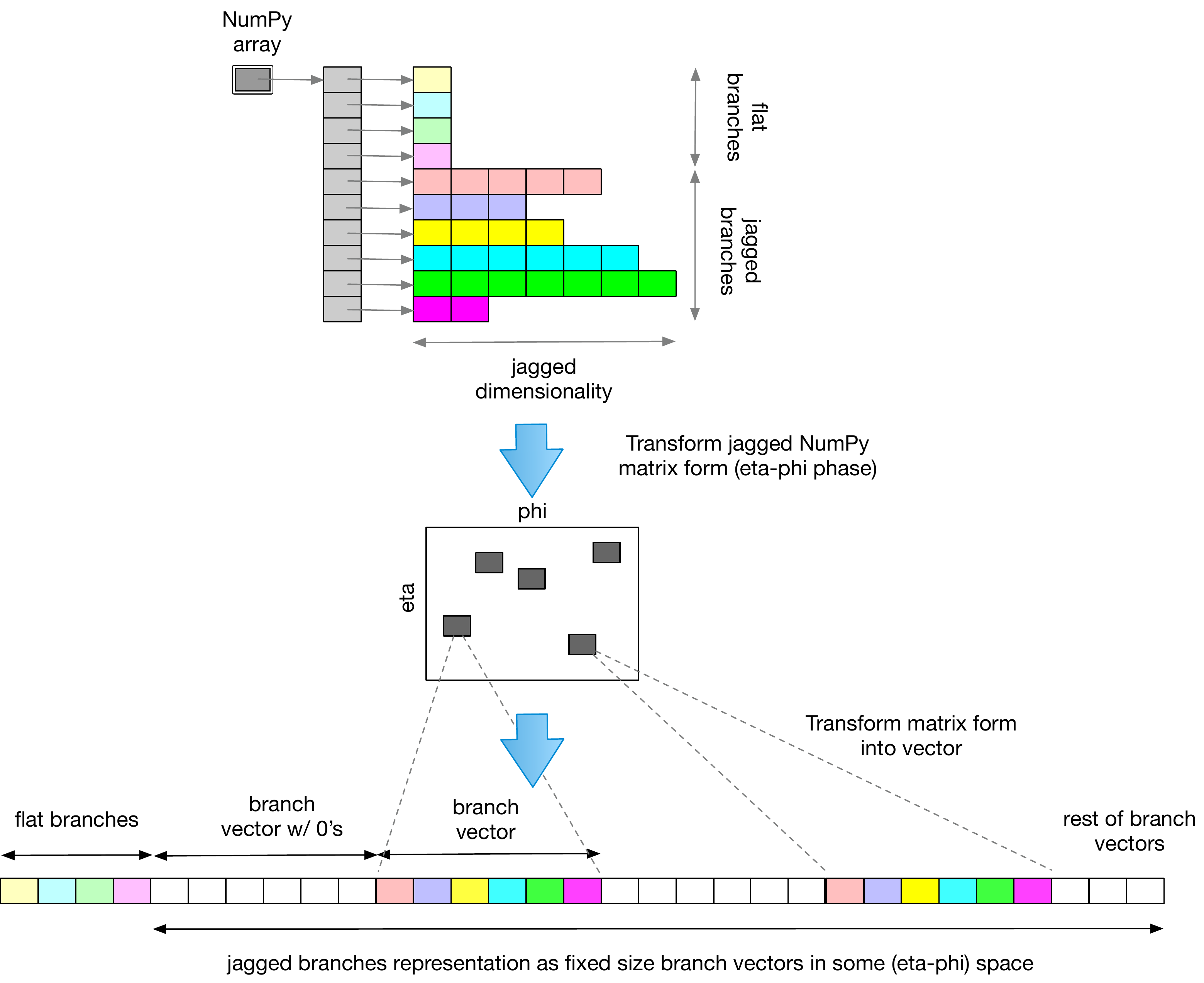}
    \caption{A matrix representation of Jagged Array into certain phase space, e.g. eta-phi.}
\label{fig:JaggedArray2Matrix}
\end{figure}

This layer is required to encapsulate HEP data and present it into ML to be
used by the application. The main obstacle here is usage of non-flat representation of HEP data in ML
frameworks. In particular, the ROOT data-format can be
represented in so called Jagged
Arrays\footnote{Jagged Array is an array of arrays of which the member arrays
can be of different sizes.}, see Fig. \ref{fig:JaggedArray}.
The HEP tree-based data representation is optimized for data storage but it is not directly
suitable for ML frameworks. Therefore a certain data transformation is required to
feed tree-based data structures into ML framework as flat data structure.
We explored two possible transformation: a vector representation with padded values,
see Fig. \ref{fig:JaggedArray2Vector}, and matrix representation into one
of the multiple phase spaces, see Fig. \ref{fig:JaggedArray2Matrix}.

The idea of the vector representation approach is to identify a dimensionality
of Jagged Array attributes in a vector via one time pass across the data, and
the subsequent composition of the final vector with sufficient allocation for
Jagged Array attribute values based on their
dimensionality.  If a certain event will have Jagged Array attribute shorter
then its dimensionality a padded values can be used. For instance, a physics
event is composed by a set of particles.  A priori we may not know how many particles
can be created in an event, and therefore we don't know how much space we need
to allocate for particle attributes even though their attributes have a fixed size, e.g.
particle momentum values can be represented by three numerical values ($p_x$,
$p_y$ and $p_z$). However, knowing the distributions of the particles in all
events of certain physics dataset can allow us to choose the dimensionality of their Jagged Array
attributes. For instance, we can run MC process and identify how many electrons
per even we may have. A maximum number of electrons in this distribution will
represent a dimensionality for corresponding Jagged Array attributes. Using
these dimensionality numbers we can represent an event as a flat vector of
certain size. The allocated values of Jagged Array attributes will vary event
by event where extra slots of Jagged Array attributes will be filled with
pre-defined pad values, e.g. NaN\footnote{Since all numerical values can be
used, e.g. in case of an angle distribution we may have negative, positive and
zero values, the only choice for padded values we have will be NaN.}.
Additionally, the one time pass across a series of events can be used to
determine the min, max, and mean values of jagged array attributes which can be
later used for normalization purposes.

The matrix representation of Jagged Array, see Fig. \ref{fig:JaggedArray2Matrix},
can use certain phase space if it is present in a dataset. For example, the spatial
coordinates or attribute components are often part of HEP datasets, and
therefore can be used for Jagged Array mappings.  This approach can resolve the
ambiguity of vector representation (in terms of dimensionality choice) but it
has its own problem with the choice of granularity of a phase space matrix. For
example, if the X-Y phase space (where X and Y refers to an arbitrary pair of
attributes) will be used in matrix presentation we don't know a cell size in
this space.
A choice of matrix granularity may introduce a collision problem with Jagged
Array attribute values, e.g. if two particles have the same phase
space values of the cell, i.e. two particles point into the same cell in
X-Y space. Such ambiguity may be easily resolved either by reducing matrix
granularity or adding other phase space, e.g. using matrices in X-Y, Y-Z and
X-Z phase spaces and concatenate them together into a final vector. But such
enhancement will increase the sparsity of the final matrix and therefore will
require more resources at the training time.

In our prototype, discussed in Sect. \ref{Prototype}, we used vector
representation with padded values and applied two pass procedure over the data.
The first pass read data streams and determined dimensionality of Jagged
Arrays along with min, max, and mean values used for normalization.
The second pass was used for reading and transforming data from
the streaming layer to the underlying ML framework.

In Neural Network models it is natural to assign padded NaN values to zeros
since they are used in the multiplication operations between input values and
weight matrix elements. But knowledge of locations of padded values in vector
representation approach may be valuable in certain circumstances. For instance,
when training AutoEncoder networks the knowledge of locations of padded values in
input vector can be used at a decoding phase.  Therefore our initial
implementation of vector representation, discussed in Sect. \ref{Prototype},
used additional mask vector to preserve the knowledge of padded values locations.

\subsection{Data Inference Layer}
A choice of a data inference layer should be driven by the usage of underlying
technology, e.g. ML framework. It can be either tightly integrated
with application frameworks (both CMS and ATLAS experiments followed
this approach in their CMSSW-DNN \cite{CMSSWDNN} and LTNN \cite{ATLASLNN}
solutions) or it can be developed as a Service (aaS) solution. The former
has the advantage of reducing latency of the inference step per processing
event, but later can be easily generalized and become independent from
the internal infrastructure. As such, it can be easily integrated into
cloud platforms, be used as repository of pre-trained models, and
serve models across experiment boundaries. We decided to implement the latter
solution via TensorFlow as a Service (TFaaS) architecture, see \cite{TFaaS}.

We evaluated several ML frameworks and decided to use TensorFlow \cite{TF}
graphs for the inference phase. The TF model represents a computational graph in a
static form, i.e.  the mathematical computations, graph edges and data flow are
well-defined at run time. Reading TF model can be done in different programming
languages due to support of APIs provided by TF library. Moreover, the TF graphs
are very well optimized for GPUs and TPUs. We chose the Go programming language to
implement the Tensor Flow as a Service (TFaaS) \cite{TFaaS} part of MLaaS
framework based on the following factors: the Go language natively supports
concurrency via goroutines and channels, it is the language developed and used
by Google and very well integrated with TF library, it 
provides a final static executable which significantly simplifies its deployment
on premises and to various (cloud) service providers.  We also opted out in
favor of REST interface where clients may upload their TF models to the TFaaS
server and use it for their inference needs via the same interface. Both Python
and C++ clients were developed on top of the REST APIs (end-points) and other
clients can be easily developed thanks to HTTP protocol used by the TFaaS Go
RESTful implementation. 

We performed several benchmarks using TFaaS server running on CentOS 7 Linux,
16 cores, 30GB of RAM.  The benchmarks were done in two modes: using 1000 calls
with 100 concurrent clients and 5000 calls with 200 concurrent clients.  We
tested both JSON and ProtoBuffer data format while sending and fetching the
data to/from TFaaS server. In both scenarios we achieved a throughput of $\sim
500$ req/sec.  These numbers were obtained with serving mid-size pre-trained
model which consists of 1024x1024 hidden layers.

Even though a single TFaaS server may not be as efficient as an integrated solution it
can be easily horizontally scaled, e.g. using kubernetes or other cluster
solutions, and may provide desire throughput for concurrent clients. It also
decouples application layer/framework from the inference phase which can be easily
integrated into any existing infrastructure by using HTTP protocol to TFaaS
server for inference results. Also, the TFaaS can be used as a repository of
pre-trained model which can be easily shared across experiment boundaries or
domains. For instance, the current implementation of TFaaS allows visual
inspection of uploaded models, versioning, tagging, etc. A simple search engine
can be put on top of TFaaS with little effort. For full list of planned improvements
see Sect. \ref{Improvements}.

\subsection{Proof-of-concept prototype}\label{Prototype}
\begin{figure}
\centering
    \includegraphics[width=0.8\textwidth]{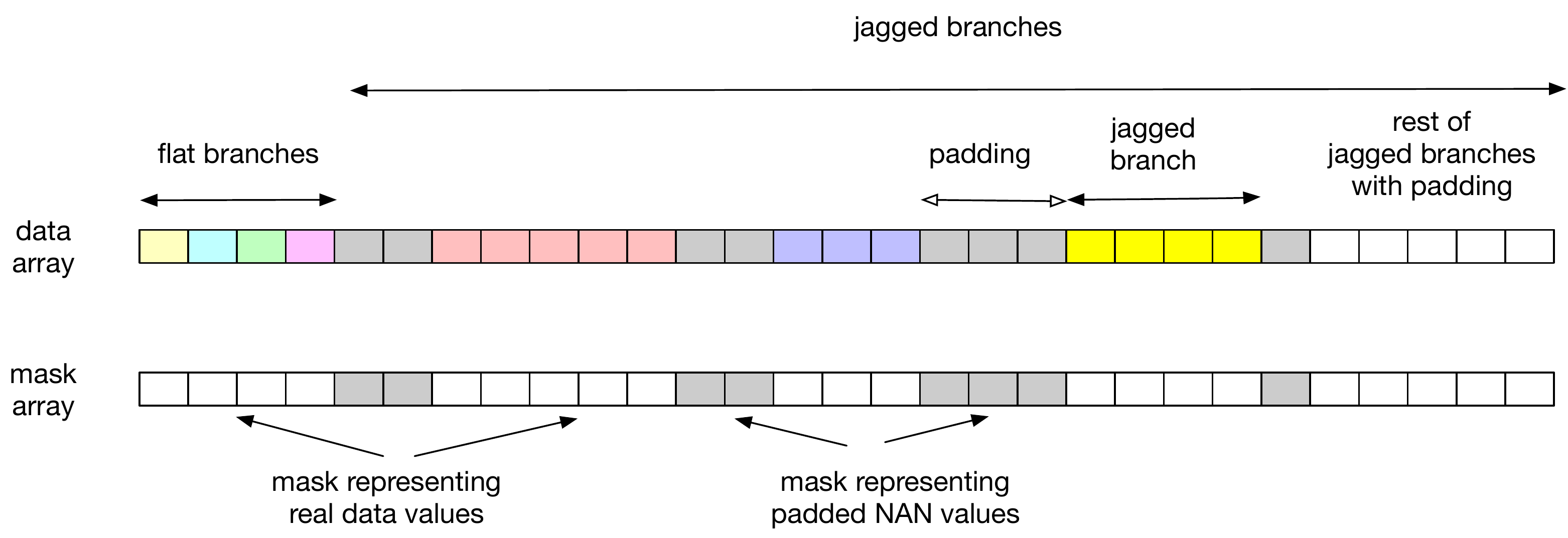}
    \caption{A vector representation of Jagged Array along with corresponding mask vector.}
\label{fig:JaggedArrayMask}
\end{figure}

When all layers of the MLaaS framework were developed, we composed a working
prototype of the system by using ROOT files accessible through XrootD servers.
The data were read by 1000 event batches, where single batch was approximately
4MB in size. Each batch was fed into both Tensor Flow (implemented via Keras
framework) and PyTorch models. The Data Generator representing
data streaming layer yields a vector representation of Jagged Array ROOT data
structures along with mask vector representing positions of padded values,
see Fig. \ref{fig:JaggedArrayMask}, into corresponding model.
This was done to avoid misinterpretation of real values of attributes from
padded values. This mask vector was used in both models to cast NaN values to
zeros. We tested this prototype on a local machine as well as successfully deploying
it on the GPU node.

The implementation of data streaming and data training layers was done in
python.  The workflow consisted of running python scripts for reading the data,
training ML models, and uploading them into TFaaS server via HTTP protocol. The
prediction was served to python, C++, and curl clients. The further details of this
proof-of-concept prototype can be found in the MLaaS4HEP github repository \cite{MLaaS4HEP}.

\section{Future directions}\label{Improvements}
We foresee that MLaaS approach can be widely applicable in HEP. As such,
further improvements will be necessary to achieve and implement.

\subsection{Data Streaming Layer}
In a Data Streaming Layer we plan to introduce proper data shuffling.
It should be done carefully when reading data from multiple remote ROOT files.
Current implementation of MLaaS reads data sequentially from file to file
and feeds the data batches directly to ML framework. In order to implement
proper data shuffling a reading parallelism should be introduced into MLaaS framework.
We also need to look at further optimization of the streaming layer to
achieve better throughput from remote data-providers.

\subsection{Data Training Layer}
The current landscape of ML framework is changing rapidly, and we should be
adapting MLaaS to existing and future ML framework and innovations. For instance, Open Network
Exchange Format \cite{onnxai} open up a door to migration of models from one framework
into another. So far we are working on automatic transformation of PyTorch \cite{PyTorch} and fast.ai
\cite{fastai} models into TensorFlow which is used by the TFaaS service.

As discussed in Sect. \ref{Training} there are different approaches
to feed Jagged Array into ML framework and R\&D in this direction is in progress.
For instance, for AutoEncoder (AE) models the vector representation with padded values
should always keep around a cast vector since AE model transform input vector into
an internal dense representation and then should decode it back into original
representation. The latter transformation can use cast vector to assign back
the padded values, and if necessary convert vector representation of the data
back to Jagged Array or ROOT TTree data-structures.

\subsection{Data Inference Layer}
On the inference side (TFaaS) we plan to extend the "aaS" part to become a
repository of uploaded models. As such, several functionalities should be added, such as
search capabilities, extended model tagging, and versioning. It can be easily achieved
by adding proper meta-data description of uploaded models and storing it
into a back-end database for later look-up, indexing and versioning.

\subsection{MLaaS services}
The proposed architecture allows to develop and deploy training and inference
layers as independent MLaaS services where separate resource providers can be
used and dynamically scaled if necessary, e.g.  GPUs/TPUs can be provisioned on
demand using commercial cloud(s) for training purposes of specific models,
while inference TFaaS service can reside at CERN premises.  For instance, the
continuous training of complex DL models would be possible when data produced
by the experiment will be placed on the GRID sites, and the training MLaaS
service will receive a set of notifications about newly available data, and
re-train specific model(s). When new model is ready it can be easily pushed to
TFaaS and be available for end-users immediately without any intervention on
the existing infrastructure. The TFaaS can be further optimized to
use FPGAs to speed up the inference phase.  We foresee that such
approach may be more flexible and cost effective for HEP experiments in HL-LHC
era.  As such, we plan to perform additional R\&D studies in this direction and
evaluate further MLaaS services using available resources.

\section{Summary}
In this paper we presented a novel Machine Learning as a Service approach to
training ML models using native ROOT format of HEP data. It consists of three
layers: data streaming, training, and inference layers, which were implemented
as independent components. The data streaming layer relies on the uproot
library for reading data from ROOT files (local or remote) and yielding NumPy
(Jagged) arrays upstream. The data training layer transforms the input Jagged
Array portion of the data into vector representation and passes it into ML
framework of user choice.  Finally, the inference layer was implemented as an
independent service (TFaaS) to serve a TensorFlow models via HTTP interface.
Such flexible architecture allows to perform ML training over HEP ROOT data
without physically downloading data into a local storage. It reads and
transforms ROOT Tree data representation (Jagged Array) into intermediate flat
data format suitable as an input for underlying ML framework.  A prototype
proof-of-concept system was developed to demonstrate MLaaS capabilities to read
arbitrary size datasets, and potentially allow to train HEP ML models over
large datasets at any scale.

\begin{acknowledgement}
This work was done as a part of CMS experiment R\&D program. I would like to
thank Jim Pivarski for his numerous and helpful discussions and hard work
on uproot (and many other) packages which open up a possibility to
implement MLaaS.
\end{acknowledgement}

%

\begin{thebibliography}{}
%

\bibitem{MLCWP}
    K. Albertsson et. al,
        {\it Machine Learning in High Energy Physics Community White Paper},
    https://arxiv.org/abs/1807.02876

\bibitem{ROOT}
    A modular scientific toolkit used in HEP for analysis and as a data-storage format,
        https://root.cern.ch/

\bibitem{kaggleATLAS}
    Higgs Boson Machine Learning Challenge used by ATLAS experiment to
        identify Higgs boson, https://www.kaggle.com/c/higgs-boson

\bibitem{kaggletracking}
    High Energy Physics particle tracking in CERN detectors, \\
    https://www.kaggle.com/c/trackml-particle-identification

\bibitem{MLaaScomparison}
    Y. Yao et. al, 
        {\it Complexity vs. performance: empirical analysis of machine learning as a service}
    Proceeding of the 2017 Internet Measurement Conference.
    ISBN: 978-1-4503-5118-8, pages 384-397, doi:10.1145/3131365.3131372

\bibitem{hls4ml}
    A package for machine learning inference in FPGAs, \\
    https://hls-fpga-machine-learning.github.io/hls4ml

\bibitem{SonicCMS}
    Services for Optimal Network Inference on Coprocessors, \\
    https://github.com/hls-fpga-machine-learning/SonicCMS

\bibitem{uproot}
    DIANA-HEP Scikit-hep uproot library, {\it Minimalist ROOT I/O in pure Python and Numpy},
    https://github.com/scikit-hep/uproot

\bibitem{DIANAHEP}
    An umbreally organization for brining state-of-the art for HEP experiements, \\
    http://diana-hep.org/

\bibitem{NumPy}
    Scientific package to represent data as multi-dimentional arrays, \\
    http://www.numpy.org/

\bibitem{XrootD}
    A high performance, scalable fault tolerant access to data repositories of many kinds,
    http://xrootd.org/

\bibitem{CMSSWDNN}
    DNN/TensorFlow interface for CMSSW, \\
        https://github.com/mharrend/CMSSW-DNN

\bibitem{ATLASLNN}
    ATALAS Lighteight Trained Neural Network, doi:10.5281/zenodo.597221, \\
    https://github.com/lwtnn/lwtnn

\bibitem{ROOT-IO}
    B. Bockelman, Z. Zhang, J. Pivarski, {\it Optimizing ROOT I/O For Analysis}, \\
    https://arxiv.org/abs/1711.02659

\bibitem{TFaaS}
    V. Kuznetsov, {\it TensorFlow as a Service} doi:10.5281/zenodo.1308048, \\
    http://github.com/vkuznet/TFaaS

\bibitem{TF}
    Tensor Flow AI library,
        http://www.tensorflow.org

\bibitem{onnxai}
    Open Neural Network Exchange format,
        http://www.onnx.ai

\bibitem{PyTorch}
    PyTorch AI library,
    https://www.pytorch.org

\bibitem{fastai}
    Fast AI library,
    https://www.fast.ai/

\bibitem{MLaaS4HEP}
    V. Kuznetsov, {\it MLaaS4HEP is set of MLaaS components for HEP}, \\
    doi:10.5281/zenodo.1481781, https://github.com/vkuznet/MLaaS4HEP

\end{thebibliography}
%
%

\end{document}